\documentclass[%
 reprint,
superscriptaddress,
 amsmath,amssymb,
 aps,
]{revtex4-2}

\usepackage{bm}%
\usepackage{float}
\usepackage{graphicx}% Include figure files
\usepackage{dcolumn}% Align table columns on decimal point
\usepackage{bm}%
\usepackage{hyperref}
\usepackage{amsmath}
\usepackage{amsfonts}
\usepackage{mathtools}
\usepackage{fixme}
\usepackage{color}
\usepackage{wasysym}

\begin{document}

%\preprint{APS/123-QED}

\title{Phase and amplitude modes in the anisotropic Dicke model with matter interactions}

\author{Ricardo Herrera Romero}
\affiliation{Departamento de F\'isica, Universidad Aut\'onoma Metropolitana-Iztapalapa, Av. Ferrocarril San Rafael Atlixco 186, C.P. 09310, CDMX, Mexico.}
\author{Miguel Angel Bastarrachea-Magnani}%
\email{bastarrachea@xanum.uam.mx}
\affiliation{Departamento de F\'isica, Universidad Aut\'onoma Metropolitana-Iztapalapa, Av. Ferrocarril San Rafael Atlixco 186, C.P. 09310, CDMX, Mexico.}

\begin{abstract}
Phase and amplitude modes are emergent phenomena that manifest across diverse physical systems, from condensed matter and particle physics to quantum optics. Also called polariton modes, we study their behavior in an anisotropic Dicke model that includes collective matter interactions. We study the low-lying spectrum in the thermodynamic limit via the Holstein-Primakoff transformation and contrast the results with the semiclassical energy surface obtained via coherent states. We also explore the geometric phase for both boson and spin contours in the parameter space as a function of the phases in the system. We unveil novel phenomena due to the unique critical features provided by the interplay between the anisotropy and matter interactions. We expect our results to serve for the observation of phase and amplitude modes in current quantum information platforms.
\end{abstract}

\maketitle

%%%%%%%%%%%%%%%%%%%%%%%%%%%%%%%%%%%%%%%%%%
%%%%%%%%%%%%%%%%%%%%%%%%%%%%%%%%%%%%%%%%%%

%%%%%%%%%%%%%%%%%%%%%%%%%%%%%%%%%%%%%%%%%%
%%%%%%%%%%%%%%%%%%%%%%%%%%%%%%%%%%%%%%%%%%
\section{Introduction}
\label{sec:1}
%%%%%%%%%%%%%%%%%%%%%%%%%%%%%%%%%%%%%%%%%%
%%%%%%%%%%%%%%%%%%%%%%%%%%%%%%%%%%%%%%%%%%

%%%%%Amplitude and phase modes.
Amplitude and phase modes are collective excitations of an order parameter usually studied in connection to a continuous symmetry breaking. For the $U(1)$ symmetry, they are known as the {\it massive} Anderson-Higgs (AH)~\cite{Anderson1963,Higgs1964} and the {\it massless} Nambu-Goldstone (NG) modes~\cite{Goldstone1961,Nambu1961}, respectively, and are present in several systems, from condensed matter~\cite{Littlewood1981,Littlewood1982,Varma2002,Pekker2015,Measson2014}, quantum antiferromagnets~\cite{Chubukov1994,Podolsky2012,Ruegg2008} to particle physics~\cite{Englert1964,Guralnik1964,Bernstein1974}. Phase and amplitude modes help to describe critical phenomena such as quantum phase transitions (QPT)~\cite{Goldstone1961}, i.e., a parametric sudden change of the ground-state’s properties of a quantum system~\cite{Sachdev1999,Carr2010}. The paradigmatic case is the quartic potential $V(\alpha)=r\alpha^{2}+g\alpha^{4}$ expressed in terms of the scalar complex parameter $\alpha$, that possesses the $U(1)$ continuous symmetry~\cite{Sachdev1999}. The parameter space is separated into two phases: the normal or non-ordered phase ($r,g>0$), where the potential has the form of a single well with a minimum at $\alpha=0$, and the ordered phase ($r<0$), where the potential has a Mexican hat form. The ground state is a degeneration of states with $|\alpha|>0$. In the ordered phase, fluctuations of the parameter quantify the two possible excitations: 1) the phase mode (NG), which occurs at constant amplitude and has vanishing excitation energy, thus massless; 2) the amplitude mode (AH), which has constant phase and possess finite excitation energy, thus being massive.
%%%%%Experimental observation
In general, both modes are hard to observe experimentally. The NG is particularly sensitive to symmetry deviations, as it gains finite excitation energy~\cite{Baksic2014,YiXiang2013}. Likewise, the AH mode is challenging to observe because of its massive character. Ultracold atoms setups are particularly suitable for their observation and control due to their high tunability~\cite{Bissbort2011,Endres2012,YiXiang2013,Leonard2017,Chiacchio2018,Schuster2020}.

%%%%%Dicke Hamiltonian.
A standard example of a mean-field quantum phase transition (QPT) is the superradiant-to-normal transition of the Dicke Hamiltonian~\cite{Hepp1973,Wang1973,Larson2017}. This model represents a collection of atoms, each simplified to a two-energy level transition, interacting with a single-mode radiation field within a cavity~\cite{Dicke1954}. When the light-matter coupling strength exceeds a critical threshold, the QPT is indicated by a non-zero photon number expectation value in the thermodynamic limit. The Dicke Hamiltonian provides a comprehensive framework for describing spin-boson interactions by capturing the collective behavior of a set of two level systems or qubits in both equilibrium and non-equilibrium setups~\cite{Garraway2011,Kirton2019,LeBoite2020,Larson2021}. Its algebraic simplicity has allowed the investigation of fundamental topics such as quantum chaos~\cite{Lambert2004,Brandes2005,Vidal2006,Villasenor2023}, the quantum-classical correspondence~\cite{Deaguiar1991,Deaguiar1992,Furuya1992,Bastarrachea2015}, Excited-State Quantum Phase Transition (ESQPT)~\cite{PerezFernandez2011,Stransky2014,Stransky2015,Cejnar2021}, non-equilibrium phenomena~\cite{Bastidas2012,Kirton2019,Kloc2018,Shen2020}, and the physics of the ultra-strong light-matter coupling regime~\cite{FornDiaz2019,FriskKockum2019,MarquezPeraca2020}. Thanks to this, the model has become a general framework to describe collective qubit effects in a vast array of systems including ultracold atoms~\cite{Nagy2010,Liu2011,Yuan2017}, superconducting qubits~\cite{Jaako2016,Yang2017,DeBernardis2018,Pilar2020}, nuclear physics~\cite{Auerbach11}, solid-state systems~\cite{Cong16,Hagenmuller12,Chirolli12}, and quantum dots~\cite{Scheibner07}. Also, superradiance has been realized in superconducting qubits~\cite{Blais04,Casanova10,Mezzacapo14}, cavity-assisted Raman transitions~\cite{Dimer2007,Baden14}, bose-Einstein condensates in optical lattices~\cite{Schneble2003,Baumann2010,Baumann2011,Klinder15,Keeling2010}, and unitary Fermi gases~\cite{Zhang2021,Helson2023}. It is precisely, in these setups where the amplitude and phase modes have been studied in connection to super-solid states exhibiting self-organization~\cite{Leonard2017,Mivehvar2018,Mivehvar2021}.

%%%%%Amplitude and phase modes in the Dicke and TC models
Under the rotating-wave approximation, the Dicke model becomes the Tavis-Cummings (TC) model~\cite{TC1968}, which is integrable in a classical sense, conserving the total number of excitations. This is reflected in the semi-classical limit, where a Mexican hat potential emerges in the superradiant phase~\cite{Bastarrachea2014a}. In this symmetry-broken phase, many states cluster together to form a quasi-continuous band of energies, where the phase can take any value in $[0,2\pi)$. Hence, there is no cost of energy to displace the state or change the order parameter along the potential's minima. Because of this, the TC model exhibits both a gapless phase mode and a gaped amplitude mode. Instead, the Dicke model has only a $\mathbb{Z}_{2}$ symmetry. As a result, it has a doubly degenerate spectrum in the superradiant phase corresponding to a double-well semi-classical potential~\cite{Bastarrachea2014a}. The pair of (symmetry-breaking) ground-states correspond to those at the bottom of a potential, reflecting two possible values for the phase ($0$ and $\pi$). Hence, the phase mode becomes gaped (Higgs-like) and corresponds to what has been called a roton-type mode~\cite{Emary2003a,Emary2003b,Mottl2012} in analogy to the roton excitations in superfluid helium. The phase and amplitude modes in the Dicke model are called {\it polariton modes}~\cite{Eastham2001}, as well. A polariton is a hybrid quantum state emerging from the strong light-matter interactions~\cite{Hopfield1958}. Because the phase and amplitude modes in the Dicke model represent the collective light-matter quantum state at low energy, they correspond to effective upper- and lower-polaritons~\cite{Carusotto2013,Ciuti2005}. The differences between the Dicke and TC models can be identified by studying the gap between the ground and first excited states. Its has been studied for finite-size~\cite{Bastarrachea2011,YiXiang2013}, and in the thermodynamical limit~\cite{Vidal2006,Liu2009}. Also, the phase and amplitude modes have been observed for large light-matter coupling~\cite{Mottl2012,Fan2014}. A way to do it is by exploiting their singular behavior at the superradiant QPT~\cite{Baksic2014}. The symmetry-breaking effect in the Dicke model has been measured experimentally~\cite{Larson2017,Baumann2011}, and employed for quantum sensing~\cite{Ivanov2015}. 

%%%%%Anisotropic Dicke model
The so-called anisotropic, generalized or unbalanced Dicke model~\cite{Deaguiar1992,Baksic2014,Bastarrachea2014a,Buijsman2017,Liu2017,Shapiro2020} is a useful tool to explore the parametric change of the gap and understanding the polariton modes as one passes from a continuous to a discrete symmetry, where the phase model can gain mass. There, one assumes a relative strength between the rotating (excitation conserving) and counter-rotating terms, allowing to tune the Hamiltonian between the TC and Dicke limits. Both the equilibrium~\cite{Bastarrachea2016,Kloc2017,Das2022,Das2023a} and non-equilibrium properties~\cite{Keeling2010,Bhaseen2012,Ferri2021,Sonriente2018,Sitely2020} of the anisotropic Dicke model has been studied, exhibiting a rich phase space and the presence of novel critical phenomena. Phase modes have been found in other extensions of the Dicke model, such as generalizations of the spin-boson coupling~\cite{Emary2004}, Jahn–Teller–Dicke models~\cite{Ivanov2013}, two-mode Dicke models~\cite{Hayn2011,Fan2014,Moodie2018,Palacino2021}, Jaynes-Cummings lattices~\cite{Hwang2016}, the three-level Dicke model~\cite{Cordero2021,LopezPena2021}, and also including matter-matter interactions~\cite{YiXiang2013}.

%%%%%Our contribution
In light of experimental progress in the combination of strong matter-matter interactions and ultra-strong light-matter systems, such as the recent realization of the Dicke model in a solid-state system with tunable spin-magnon interactions~\cite{Li2018,Bamba2022,MarquezPeraca2024}, in this work, we investigate the effects of collective matter interactions over the phase and amplitude or polariton modes in the anisotropic Dicke model. Collective matter interactions mediated by light introduce unique behavior in the Dicke model such as first-order QPT~\cite{Lee2004,Chen2008,Chen2010,Rodriguez2011,Zhao2017,Rodriguez2018,Yang2019}, parametric shifts over the superradiant QPT~\cite{Chen2006,Nie2009,Jaako2016}, new quantum phases~\cite{Robles2015,Rodriguez2018,Herrera2022,Liu2023}, and novel perspectives on quantum chaos~\cite{Rodriguez2018,Sinha2020}, such as the amplification of regularity-to-chaos transition~\cite{Wang2022}.

To achieve our end, we employ the standard approach of applying the Holstein-Primakoff (HP) transformation to obtain a low-energy approximation of the excitation modes around the minima~\cite{Emary2003}. Given the limitations of the HP transformation~\cite{Hirsch2013}, we interpret the results under the light of the semiclassical corresponding energy surfaces that can be obtained via coherent states~\cite{Herrera2022}. Additionally, we calculate the geometric or Berry phase around the critical points and relate its behavior to the phase and amplitude modes~\cite{Berry1984,Berry1985}. The geometric phase possess a non-analytic behavior along a QPT, so it has been employed as a signature of criticality~\cite{Carollo2005,Pachos2006,Reuter2007}. It has been observed experimentally in Heisenberg chains~\cite{Peng2010} and qubit systems~\cite{Leek2007,Zhang2017}. It has been studied for the standard Dicke model~\cite{Plastina2006,Chen2006b,Carollo2020} and extensions such as including two-photon procesess~\cite{Guerra2020}, impurity coupling~\cite{Lu2022}, and collective matter interactions in the $z$ direction~\cite{Li2013}.

%%%%%Organization
The article is organized as follows. In Sec.~\ref{sec:2}, we present the anisotropic Dicke Hamiltonian with matter interactions and details over its classical limit. Next, in  Sec.~\ref{sec:3}, we obtain exact expressions for the low-lying polariton branches using the Holstein-Primakoff approximation. In Sec.~\ref{sec:4}, we discuss the gap behavior as a function of the different phases in the system. We calculate the geometrical phase for arbitrary circulations generated by the photon number or the relative population operator in Sec.~\ref{sec:5}. Finally, in Sec.~\ref{sec:6}, we present our conclusions and perspectives.

%%%%%%%%%%%%%%%%%%%%%%%%%%%%%%%%%%%%%%%%%%
%%%%%%%%%%%%%%%%%%%%%%%%%%%%%%%%%%%%%%%%%%
\section{Anisotropic Dicke Hamiltonian with matter interactions}
\label{sec:2}
%%%%%%%%%%%%%%%%%%%%%%%%%%%%%%%%%%%%%%%%%%
%%%%%%%%%%%%%%%%%%%%%%%%%%%%%%%%%%%%%%%%%%
The anisotropic Dicke Hamiltonian including collective matter interactions is given by
\begin{gather} \label{eq:1}
        \hat{H}_{\xi}=\omega \hat{a}^{\dagger}\hat{a}+\omega_{0}\hat{J}_{z}+\\ \nonumber \frac{\gamma}{\sqrt{N}}\left[(\hat{a}\hat{J}_{+}+\hat{a}^{\dagger}\hat{J}_{-})+\xi(\hat{a}^{\dagger}\hat{J}_{+}+\hat{a}\hat{J}_{-})\right]+\\ \nonumber \frac{1}{N}\left(\eta_{x}\hat{J}_{x}^{2}+\eta_{y}\hat{J}_{y}^{2}+\eta_{z}\hat{J}^{2}_{z}\right),
\end{gather}
The first two terms denote the non-interacting parts of the Hamiltonian, where $\omega$ is the boson frequency, $\omega_{0}$ is the qubit energy splitting, $\hat{a}^{\dagger}$ ($\hat{a}$) is the creation (annihilation) boson operator, $\hat{J}_{z}$ is the relative population of excited qubits, given that $\hat{J}_{z,x,y}$ are pseudo-spin operators representing the collective degrees of freedom of the set of $N$ qubits, which follow the rules of the $\text{su(2)}$-algebra. The third term is the spin-boson interaction, with $\gamma$ the coupling strength and $\xi$ the anisotropy parameter, that allows tuning the relative strength of the counter-rotating term, so $\xi=0$ and $\xi=1$ are the TC and Dicke limits, respectively. The last term accounts for the matter interactions, where $\eta_{i}$ with $i=x,y,z$ are the couplings in each direction, whose meaning depends on the particular setup realizing the Dicke model, including Josephson dynamics~\cite{Chen2007,Rodriguez2011,Sinha2019}, atomic dipolar couplings~\cite{Joshi1991,Chen2006}, optomechanical setups~\cite{Abdel-Rady2017,Salah2018}, or interactions between superconducting qubits~\cite{Tian2010,Zhang2014,Jaako2016,DeBernardis2018,Pilar2020}. A standard approach is regarding $\eta_{z}$ as the strength of collective on-site interaction and $\eta_{x}$ ($\eta_{y}$) as a strength of the collective inter-qubit interactions. It is worth emphasizing that the relevant interacting parameters are $\Delta\eta_{z\mu}=\eta_{z}-\eta_{\mu}$, with $\mu=x,y$, given that the pseudospin length is conserved, and one can express one direction in terms of the others  $\hat{J}^{2}_{z}=j(j+1)\hat{\mathbb{I}}-\hat{J}^{2}_{x}-\hat{J}^{2}_{y}$~\cite{Herrera2022}.

The Hamiltonian in Eq.~\ref{eq:1} commutes with the squared pseudospin length operator, $[\hat{H}_{\xi},\hat{\mathbf{J}^{2}}]$, so the collective ground-state lies in the totally symmetric subspace that corresponds to $j=N/2$, and the Hilbert space of the system is effectively reduced to $N+1$ states. As a result, a classical corresponding Hamiltonian can be obtained by mapping the boson and collective pseudospin degrees of freedom to the classical limit via coherent states. To do so, one takes the value of the quantum Hamiltonian over the tensor product of Glauber $|z\rangle$ and Bloch $|w\rangle$ coherent states as trial states~\cite{Deaguiar1992,Bastarrachea2014a,Bastarrachea2014b,Bastarrachea2015,Chavez2016}, where $|0\rangle$ and $|j,-j\rangle$ are the boson and pseudo-spin vacuum states, respectively~\cite{Gilmore1990}. In canonical classical variables for the boson $(q,p)$, $z=\sqrt{j/2}\left(q+ip\right)$ and the pseudo-spin $(j_z,\phi)$, $w=\sqrt{(1+j_z)/(1-j_z)}e^{-i\phi}$ the classical corresponding Hamiltonian reads~\cite{Herrera2022}
\begin{gather}\label{eq:class}
    H_{\xi}^{(cl)}=\frac{\omega}{2}(q^{2}+p^{2})+j_{z}\left(\omega_{0}+\frac{\eta_{z}j_{z}}{2}\right)+\\ \nonumber 
    \frac{1}{2}\left(1-j^{2}_{z}\right)\left(\eta_{x}\cos^{2}\phi+\eta_{y}\sin^{2}\phi\right)+\\ \nonumber
    +\gamma\sqrt{1-j^{2}_{z}}\left[(1+\xi)q\cos\phi-(1-\xi)p\sin\phi\right].
\end{gather}

A standard method to determine quantum phases in the system is to analyze the extreme points of the classical energy surface in Eq.~\ref{eq:class}, i.e., finding fixed points of the corresponding Hamilton equations because they can be related to critical changes in the systems. Hence, absolute minima signal ground-state QPT and other extreme points to excited-state QPT~\cite{Bastarrachea2014a,Bastarrachea2016,Rodriguez2018}. In the following, we will focus only on the absolute minima of the energy surfaces, as we are interested in polariton modes. We employ the classical energy surfaces to interpret the behavior of these modes across the parameter space. We picture the surfaces in a set of variables $(u,v)=\arccos(-j_{z})(\cos\phi,\sin\phi)$ that, by using Hamilton equations to eliminate the bosonic quadratures $q$ and $p$, allows one to visualize the fixed points in the collective pseudospin space alone~\cite{Bastarrachea2014a,Herrera2022}. In this picture, the energy surface reads as
\begin{gather}
E(\xi,u,v)=\omega_{0}\sin^{2}\sqrt{u^{2}+v^{2}}\frac{1}{2\left(u^{2}+v^{2}\right)}\times \\ \nonumber 
\left[u^{2}\left(\frac{\eta_{x}}{\omega_{0}}-f_{\xi +}\right)+v^{2}\left(\frac{\eta_{y}}{\omega_{0}}-f_{\xi -}\right)\right]-\\ \omega_{0}\cos\sqrt{u^{2}+v^{2}}\nonumber\left(1-\frac{\eta_{z}}{2\omega_{0}}\cos\sqrt{u^{2}+v^{2}}\right).
\end{gather}
where $f_{\xi\pm}=\gamma^{2}/\gamma_{\xi\pm}^{c}$, and $\gamma_{\xi\pm}^{c}=\sqrt{\omega\omega_{0}}/(1\pm\xi)$, are the critical couplings for the superradiant-$(\pm)$ phases as detailed below.

The anisotropy introduces new quantum phases to the standard Dicke model that matter-matter interactions then modify. For zero matter interactions ($\eta_{I}=0$), the Dicke ($\xi=1$) and TC ($\xi=0$) models exhibit only two phases, the normal (non-ordered) and the superradiant (ordered), separated by the critical coupling $\gamma_{\xi+}^{c}$. The anisotropy modifies the energy landscape such that two different superimposed phases appear, the superradiant-$(+)$ and superradiant-$(-)$ phases, the latter with a critical coupling $\gamma_{\xi-}^{c}=$~\cite{Bastarrachea2016,Kloc2017,Cejnar2021}. These phases are also called the electric and magnetic superradiant, with a different coupling combination for the counter and counter-rotating terms~\cite{Baksic2014,Shapiro2020}. Each phase has its extreme points, but only one predominates as the ground-state, depending on which of the two terms, the rotating or the counter-rotating, is stronger. For our case of study $\xi\in[0,1]$, the superradiant-$x$ phase defines the ground-state.

In the presence of interactions ($\eta_{i}\neq0$), but in the absence of light-matter coupling ($\gamma=0$) one gets a Lipkin-Meshkov-Glick (LMG) Hamiltonian~\cite{Lipkin65,Meshkov65,Glick65}. Hence, the anisotropic Dicke Hamiltonian inherits the critical phenomena from the LMG~\cite{Dusuel04,Dusuel05,Castanos06,Heiss05,Leyvraz05,Heiss06,Ribeiro08,Engelhardt15,GarciaRamos17}. There are three major modifications to the ground-state quantum phase diagram of the anisotropic Dicke model due to matter interactions~\cite{Herrera2022}. First, interactions in the $x$ and $y$ direction produce a deformation, creating a {\it deformed} normal phase~\cite{Shapiro2020}. Second, the superradiant-$(+)$ (superradiant-$(-)$) phase transforms into the superradiant-$x$ (superradiant-$y$), and the critical coupling is modified to $\gamma_{\xi x}^{c}=(1+\Delta\eta_{zx}/\omega_{0})^{1/2}\gamma_{\xi +}^{c}$ ($\gamma_{\xi y}^{c}=(1+\Delta\eta_{zy}/\omega_{0})^{1/2}\gamma_{\xi -}^{c}$)~\cite{Herrera2022}. Finally, while $\eta_{z}$ produces an overall energy shift,
in the case of the Dicke limit ($\xi=1$), a new phase emerges,  the deformed phase, which suppresses superradiance for $\Delta\eta_{zy}\geq 1$ and is the source of a first-order QPT for arbitrary coupling~\cite{Rodriguez2018}.  

%%%%%%%%%%%%%%%%%%%%%%%%%%%%%%%%%%%%%%%%%%
%%%%%%%%%%%%%%%%%%%%%%%%%%%%%%%%%%%%%%%%%%
\section{Phase and amplitude modes in the anisotropic Dicke model}
\label{sec:3}
%%%%%%%%%%%%%%%%%%%%%%%%%%%%%%%%%%%%%%%%%%
%%%%%%%%%%%%%%%%%%%%%%%%%%%%%%%%%%%%%%%%%%

The Holstein-Primakoff transformation (HPT)~\cite{Holstein1940} provides a quadratic approximation to the lower energy modes of Eq.~\ref{eq:1} in the thermodynamic limit ($N\rightarrow\infty$). After diagonalizing the quadratic, approximated Hamiltonian, the resulting branches correspond to the polariton modes or mean-field low-lying excitations of the photon-matter quantum superposition~\cite{Emary2003b,Baksic2014}. The HPT reads
\begin{gather}\label{eq:2}
\hat{J}_{z}=\hat{b}^{\dagger}\hat{b}-j,\,\,\,
\\ \nonumber
\hat{J}_{+}=\sqrt{2j}\,\hat{b}^{\dagger}
\sqrt{1-\frac{\hat{b}^{\dagger}\hat{b}}{2j}},\,\,\,
%\simeq\sqrt{2j}\hat{b}^{\dagger},
\\ \nonumber
\hat{J}_{-}=\sqrt{2j}\,
\sqrt{1-\frac{\hat{b}^{\dagger}\hat{b}}{2j}}\,\hat{b}.
%\simeq\sqrt{2j}\hat{b},
\end{gather}
where $\hat{b}$ ($\hat{b}^{\dagger}$) is an annhilation (creation) boson operator, such that
$[\hat{b},\hat{b}^{\dagger}]=\mathbb{I}$. First, by substituting Eqs.~\ref{eq:2} one obtains
%%%
\begin{gather}
    \nonumber
    \hat{H} = \omega \hat{a}^{\dagger}\hat{a} + \omega_{0}\left(\hat{b}^{\dagger}\hat{b} -j\right) \\
    %%%
    + \gamma \left[ \left( \hat{a}\hat{b}^{\dagger}\sqrt{1-\frac{\hat{b}^{\dagger}\hat{b}}{2j}} + \hat{a}^{\dagger}\sqrt{1-\frac{\hat{b}^{\dagger}\hat{b}}{2j}}\hat{b} \right) + \right.
    \\ \nonumber 
    \left.\xi\left( \hat{a}^{\dagger}\hat{b}^{\dagger}\sqrt{1-\frac{\hat{b}^{\dagger}\hat{b}}{2j}} + \hat{a}\sqrt{1-\frac{\hat{b}^{\dagger}\hat{b}}{2j}}\hat{b} \right) \right] \\
    \nonumber
    %%%
    \frac{1}{4}\left[\eta_{x}\left(\hat{b}^{\dagger}\sqrt{1-\frac{\hat{b}^{\dagger}\hat{b}}{2j}} + \sqrt{1-\frac{\hat{b}^{\dagger}\hat{b}}{2j}}\hat{b}\right)^{2}-\right.
    \\ \nonumber \left.\eta_{y}\left(\hat{b}^{\dagger}\sqrt{1-\frac{\hat{b}^{\dagger}\hat{b}}{2j}}-\sqrt{1-\frac{\hat{b}^{\dagger}\hat{b}}{2j}}\hat{b}\right)^{2} \right] + \frac{\eta_{z}}{2j}\left( \hat{b}^{\dagger}\hat{b}-j\right)^{2}
\end{gather}
Next, to study the excitation modes we employ a mean field approximation by displacing both $\hat{a}$ and $\hat{b}$ bosonic operators 
\begin{gather}
    \hat{a}^{\dagger} \to \hat{c}^{\dagger} + \alpha\sqrt{2j}, \;\;\; \text{and} \;\;\;
    \hat{b}^{\dagger} \to \hat{d}^{\dagger} -\beta\sqrt{2j}. 
\end{gather}
where $\alpha, \beta \in  \mathbb{C}$ are scaled displacements. The square roots in the HPT become
%%%
\begin{gather}
\label{kxi}
    \sqrt{1-\frac{\hat{b}^{\dagger}\hat{b}}{2j}} = \\ \nonumber  \sqrt{1-\frac{ \hat{d}^{\dagger}\hat{d} - \beta\sqrt{2j}(\hat{d}^{\dagger} + \hat{d}) + 2j\beta^{2}}{2j}} = \sqrt{k}\sqrt{\xi_{0}}, \\
    \nonumber 
    k = 1 - \beta^{2}, \;\;\;  \sqrt{\xi_{0}} = \sqrt{1-\frac{\hat{d}^{\dagger}\hat{d} - \beta\sqrt{2j}(\hat{d}^{\dagger} + \hat{d})+2j\beta^{2}}{2jk}}
\end{gather}
%%%
Next, we make use of the thermodynamic limit $N\to \infty$. Here, the customary assumption is that $\langle\hat{b}^{\dagger}\hat{b}\rangle\ll 2j$, so $\sqrt{1-\hat{b}^{\dagger}\hat{b}/2j}\simeq 1-\hat{b}^{\dagger}\hat{b}$. In particular, 
\begin{gather}
\label{taylor apprx}
    \sqrt{\xi_{0}}\simeq
    1 - \frac{\hat{d}^{\dagger}\hat{d} - \beta\sqrt{2j}(\hat{d}^{\dagger} \hat{d})}{4jk}-\frac{\beta^{2}(\hat{d}^{\dagger}+\hat{d})^{2}}{16jk^{2}}
\end{gather}
By keeping only terms up to quadratic order, the Hamiltonian becomes
%%%
\begin{widetext}
\begin{gather} \label{eq:hc}
    \hat{H} = \omega \hat{c}^{\dagger}\hat{c} + \left[ \omega_{0} + \gamma\frac{\alpha\beta}{\sqrt{k}}(1+\xi) + \eta_{z}(2\beta^{2}-1) \right]\hat{d}^{\dagger}\hat{d} + \left[ \omega\alpha - \gamma\sqrt{k}\beta(1+\xi) \right]\sqrt{2j}(\hat{c}^{\dagger} + \hat{c}) \nonumber \\
    %%%
    +\left[ -\omega_{0}\beta + \gamma\alpha\left( \frac{k-\beta^{2}}{\sqrt{k}}\right)(1+\xi) - \eta_{x}k\beta\left(1-\frac{\beta^{2}}{k}\right)-\eta_{z}(2\beta^{2}-1)\beta\right]\sqrt{2j}(\hat{d}^{\dagger} + \hat{d}) \nonumber \\
    %%%
    +\left[ \gamma\frac{\alpha\beta}{4\sqrt{k}k}(2-\beta^{2})(1+\xi) + \frac{k}{4}\eta_{x}\left(1-\frac{\beta^{2}}{k}\right)^2 + \eta_{z}\beta^{2} \right](\hat{d}^{\dagger} + \hat{d})^{2} - \eta_{y}\frac{k}{4}(\hat{d}^{\dagger}-\hat{d})^{2}  \nonumber \\
    %%%
    -\gamma\left[ \frac{\beta^{2}}{2\sqrt{k}}(1+\xi) \right](\hat{c}^{\dagger} + \hat{c})(\hat{d}^{\dagger} + \hat{d}) + \gamma\sqrt{\xi_{0}}\left[  (\hat{c}d^{\dagger} +\hat{c}^{\dagger}\hat{d}) + \xi(\hat{c}^{\dagger}\hat{d}^{\dagger} + \hat{c}\hat{d}) \right] \nonumber \\
    %%%
    \label{lineal hamiltonian}
    + \left[\omega\alpha^{2} + \omega_{0}\beta^{2} - \frac{\omega_{0}}{2} - \gamma\sqrt{k}2\alpha\beta(1+\xi) + \eta_{x}k\beta^{2} + \eta_{z}\left(\beta^{2}-\frac{1}{2}\right)^{2} \right]2j -\gamma\frac{\alpha\beta}{2\sqrt{k}}(1+\xi). 
\end{gather}
\end{widetext}

%%%%%%%%%%%%%%%%%%%%%%%%%%%%%%%%%%%%%%%%%%
\subsection{Deformed Normal Phases}
%%%%%%%%%%%%%%%%%%%%%%%%%%%%%%%%%%%%%%%%%%

The deformed normal phases are characterized by a zero expectation value of photon and matter excitations. Thus, by setting $\alpha,\beta\rightarrow 0$, we obtain
\begin{gather}
 \nonumber
    \hat{H} = \omega \hat{c}^{\dagger}\hat{c} + \omega_{0}\left(\hat{d}^{\dagger}\hat{d} -j\right) \\ \nonumber 
    %%%
    + \gamma \left[ \left( \hat{c}\hat{d}^{\dagger} + \hat{c}^{\dagger}\hat{d} \right) + \xi\left( \hat{c}^{\dagger}\hat{d}^{\dagger} + \hat{c}\hat{d} \right) \right] 
    +\\ \nonumber 
    %%%
    \frac{1}{4}\left[\eta_{x}\left(\hat{d}^{\dagger} + \hat{d}\right)^{2}-\eta_{y}\left(\hat{d}^{\dagger}-\hat{d}\right)^{2}\right]
    + \frac{\eta_{z}}{2j}\left( \hat{d}^{\dagger}\hat{d}-j\right)^{2}.
\end{gather}
%%%
Next, one writes the Hamiltonian in terms of the boson quadratures~\cite{Emary2003}
%%%
\begin{gather}\label{eq:mf}
    \hat{c} = \sqrt{\frac{\omega}{2}}\left( \hat{x} + \frac{i}{\omega}\hat{p}_{x} \right),\;\;\; %\hat{a}^{\dagger} = \sqrt{\frac{\omega}{2}}\left( \hat{x} -\frac{i}{\omega}\hat{p}_{x} \right),
    \hat{d} = \sqrt{\frac{\omega_{0}}{2}}\left( \hat{y} + \frac{i}{\omega_{0}}\hat{p}_{y} \right),
    %\hat{b}^{\dagger} = \sqrt{\frac{\omega_{0}}{2}}\left( \hat{y} -\frac{i}{\omega_{0}}\hat{p}_{y} \right).  
\end{gather}
%%%
By substituting Eqs.~\ref{eq:mf}, the Hamiltonian is now represented as follows:
%%%
\begin{gather} \label{eq:hm1}
    \hat{H} = \frac{1}{2}\left[ \omega^{2}\hat{x}^2 + \omega_{zx}^{2} \hat{y}^{2}  + \hat{p}_{x}^{2} + \omega_{zy}^{2} \hat{p}_{y}^{2} +\right.
    \\ \nonumber 
    \left.\frac{\gamma}{\gamma_{\xi +}}2\omega\omega_{0}\hat{x}\hat{y} + \frac{\gamma}{\gamma_{\xi -}}2\hat{p}_{x}\hat{p}_{y} -\epsilon_{0}\right].
\end{gather}
where 
\begin{gather}
\epsilon_{0}=\omega + \omega_{0}\left(1 - \frac{\eta_{z}}{\omega_{0}}\right) + 2j\omega_{0}\left(1-\frac{\eta_{z}}{2\omega_{0}}\right), 
\end{gather}
%%%
We considering the following dimensionless frequencies containing the matter interaction strengths
%%%
\begin{gather}
    \tilde{\omega}_{zx} = \sqrt{1 - \frac{\Delta\eta_{zx}}{\omega_{0}}}, \;\;\;
     \text{and} \;\;\;
    %%%
    \tilde{\omega}_{zy} = \sqrt{1 - \frac{\Delta\eta_{zy}}{\omega_{0}}}.
\end{gather}
Next, two rotations are performed to eliminate the cross terms $\hat{x}\hat{y}$ and $\hat{p}_{x}\hat{p}_{y}$. We will apply a rotation with angle $\theta_{1}$ for the variables $\hat{x}$ and $\hat{y}$, and another with angle $\theta_{2}$ for $\hat{p}{x}$ and $\hat{p}{y}$.
%%%
\begin{gather}
\left(
\begin{array}{c}
\hat{x} \\
\hat{y}  
\end{array}
\right)
\left(
\begin{array}{cc}
\cos{\theta_{1}} & \sin{\theta_{1}} \\
-\sin{\theta_{1}} & \cos{\theta_{1}}
\end{array}
\right)
\left(
\begin{array}{c}
\hat{q}_{1} \\
\hat{q}_{2}  
\end{array}
\right),\;\;\; \\ \nonumber 
\left(
\begin{array}{c}
\hat{p}_{x} \\
\hat{p}_{y}  
\end{array}
\right)
\left(
\begin{array}{cc}
\cos{\theta_{2}} & \sin{\theta_{2}} \\
-\sin{\theta_{2}} & \cos{\theta_{2}}
\end{array}
\right)
\left(
\begin{array}{c}
\hat{p}_{1} \\
\hat{p}_{2}  
\end{array}
\right).
\end{gather}
%%%
The following conditions arise:
%%%
\begin{gather}
    \tan{2\theta_{1}} = 2f_{\xi x}^{1/2}\frac{\omega\omega_{0}\tilde{\omega}_{zx}}{\omega_{0}^{2}\tilde{\omega}_{zx}^{2}-\omega^{2}}, \;\;\; \\ \nonumber 
    %%% 
    \text{and} \;\;\;
    %%%
    \tan2\theta_{2} = 2f_{\xi x}^{1/2}\left(\frac{1-\xi}{1+\xi}\right)\frac{\tilde{\omega}_{zx}}{\tilde{\omega}_{zy}^{4}-1},
\end{gather}
%%%
where $f_{\xi x}=(\gamma/\gamma_{\xi x})^{2}$, and the critical coupling indicating the onset of superradiance is
%%%
\begin{gather} \label{eq:gc}
    \gamma_{\xi x}^{c} = \gamma_{\xi +}^{c}\tilde{\omega}_{zx}.
\end{gather}
Substituting these conditions yields a two-mode decoupled Hamiltonian, determining the energies $\epsilon_{1\pm}^{N}$ that represent the low-lying excitation modes.
%%%
\begin{gather}
\label{Normal Hamiltonian N pq 1}
    \hat{H} = \frac{1}{2}\left[     (\epsilon_{1-}^{\text{N}})^{2}\hat{q}_{1}^{2}  + (\epsilon_{1+}^{\text{N}})^{2}\hat{q}_{2}^{2} + \right.
    \\ \nonumber 
    \left.(\epsilon_{2-}^{\text{N}})^{2}\hat{p}_{1}^{2}  + (\epsilon_{2+}^{\text{N}})^{2}\hat{p}_{2}^{2}
     -\epsilon_{0}\right],
\end{gather}
with energies are
\begin{widetext}
\begin{gather}
    \label{eq:eps1Npm}
    \epsilon_{1\pm}^{N} = \sqrt{\frac{1}{2}\left[ (\omega^{2} + \omega_{0}^{2}\tilde{\omega}_{zx}^{2}) \pm \sqrt{(\omega^{2}-\omega_{0}^{2}\tilde{\omega}_{zx}^{2})^{2} + 4\omega^{2}\omega_{0}^{2}\tilde{\omega}^{2}_{zx}f_{\xi x} } \right]}, \\
    %%%
    \label{eq:eps2Npm}
    \epsilon_{2\pm}^{N} = \sqrt{\frac{1}{2}\left[(1+\tilde{\omega}_{zy}^{2}) \pm \sqrt{(1-\tilde{\omega}_{zy}^{2})^{2}+4\tilde{\omega}_{zy}^{2}g_{\xi y}}\right]}
\end{gather}
\end{widetext}
%%%
where $g_{\xi y}=\left(\gamma/\gamma_{\xi y}\right)^{2}$, with 
\begin{gather}
\gamma_{\xi y}=\gamma_{\xi -}^{c}\tilde{\omega}_{zy}
\end{gather}
the critical coupling of the superradiant-$y$ behavior. It is convenient to rewrite Eq.~\ref{Normal Hamiltonian N pq 1} in a second quantized form in order to obtain a decoupled Hamiltonian with new bosonic excitations $\hat{a}_{i}$ ($\hat{a}_{i}^{\dagger}$) with $i=1,2$ following $[\hat{a}_{i},\hat{a}_{i}^{\dagger}]=\mathbb{I}$. They read
%%%
\begin{gather}
    \hat{q}_{1} = \frac{(\hat{a}^{\dagger}_{1} + \hat{a}_{1})}{\sqrt{2\omega_{\text{N}_{-}}}}, \;\;\; \hat{p}_{1} = i\sqrt{\frac{\omega_{\text{N}_{-}}}{2}}(\hat{a}^{\dagger}_{1} - \hat{a}_{1}), \\
    %%%
    \hat{q}_{2} = \frac{(\hat{a}^{\dagger}_{2} + \hat{a}_{2})}{\sqrt{2\omega_{\text{N}_{+}}}}, \;\;\; \hat{p}_{2} = i\sqrt{\frac{\omega_{\text{N}_{+}}}{2}}(\hat{a}^{\dagger}_{2} - \hat{a}_{2}), 
\end{gather}
%%%
where $\omega_{\text{N}_{-}} =\epsilon_{1-}^{\text{N}}/\epsilon_{2-}^{\text{N}}$, and $\omega_{\text{N}_{+}} =\epsilon_{1+}^{\text{N}}/\epsilon_{2+}^{\text{N}}$. Thus, we obtain the quantum oscillation modes and the low-energy spectrum. It is essential to highlight that these expressions depend on the $\Delta\eta_{xz}$ and $\Delta\eta_{yz}$ parameters, allowing us to modulate the behavior of the modes both in the normal phase and in the superradiant phase as we will see later on. Finally, the Hamiltonian reads
%%%
\begin{gather}
    \hat{H} = \epsilon_{0}^{\text{N}}+ \epsilon_{-}^{\text{N}}\hat{a}_{1}^{\dagger}\hat{a}_{1} + \epsilon_{+}^{\text{N}}\hat{a}_{2}^{\dagger}\hat{a}_{2},
\end{gather}
with 
\begin{gather}
\epsilon_{0}^{\text{N}}=\frac{1}{2}\left(\epsilon_{-}^{\text{N}} + \epsilon_{+}^{\text{N}}  -\epsilon_{0} \right)
\end{gather}
The phase $\epsilon_{-}^{\text{N}}$ and amplitude $\epsilon_{+}^{\text{N}}$ modes in the deformed normal phases become
%%%
\begin{gather}
\label{Modos de vibración Normal}
    \epsilon_{-}^{\text{N}} = \epsilon_{2-}^{\text{N}}\epsilon_{1-}^{\text{N}}, \;\;\; \epsilon_{+}^{\text{N}} = \epsilon_{2+}^{\text{N}}\epsilon_{1+}^{\text{N}}.
\end{gather}

%%%%%%%%%%%%%%%%%%%%%%%%%%%%%%%%%%%%%%%%%%
\subsection{Superradiant phases}
%%%%%%%%%%%%%%%%%%%%%%%%%%%%%%%%%%%%%%%%%%

Now, we consider the case where $\alpha,\beta\neq 0$. To obtain a quadratic Hamiltonian, we need to eliminate the linear terms in $\hat{c},\hat{c}^{\dagger}$, and $\hat{d},\hat{d}^{\dagger}$ from  Eq.~\ref{eq:hc}. Hence, we obtain the following conditions determining the values of $\alpha$ and $\beta$,
%%%
\begin{gather}
\label{HP LINEAL 1}
    \omega\alpha - \gamma\sqrt{k}\beta(1+\xi) = 0, \\
    %%%
\label{HP LINEAL 2}
     -\omega_{0}\beta + \gamma\alpha\left( \frac{k-\beta^{2}}{\sqrt{k}}\right)(1+\xi) - \\ \nonumber 
     \eta_{x}k\beta\left(1-\frac{\beta^{2}}{k}\right) - \eta_{z}(2\beta^{2}-1)\beta = 0
\end{gather}
%%%
Solving the system of equations leads to
%%%
\begin{gather} 
\label{alphabeta}
    \alpha = \frac{\gamma(1+\xi)}{2\omega}\sqrt{1-\mu_{x}^{2}} \;\;\;\text{and}\;\;\; \beta = \sqrt{\frac{1}{2}\left(1-\mu_{x}\right)},
\end{gather}
%%%
where
%%%
\begin{gather*}\label{eqmux}
    \mu_{x} = \left[\tilde{\omega}_{zx}^{2}\left(f_{\xi x}-1\right)+1\right]^{-1}.
\end{gather*}
%%%
The critical coupling $\gamma_{\xi x}^{c}$ in Eq.~\ref{eq:gc} for the existence of superradiant behavior is determined by $\mu_{x}=1$, because for $\mu_{x}>1$ $\alpha$ and $\beta$ in Eq.~\ref{alphabeta} become complex. Writing the resulting Hamiltonian in terms of Eq.~\ref{alphabeta}, we have:
%%%
\begin{gather}
    \hat{H} = \omega \hat{c}^{\dagger}\hat{c} + \omega_{A}\hat{d}^{\dagger}\hat{d} 
    + \omega_{B}(\hat{d}^{\dagger} \\ \nonumber   + \hat{d})^{2} + \omega_{C}(\hat{d}^{\dagger}-\hat{d})^{2} 
    +\omega_{D}(\hat{c}^{\dagger} + \hat{c})(\hat{d}^{\dagger} + \hat{d}) \nonumber \\ \nonumber \label{Hamiltonian frequencies}
    %%%
    + \omega_{E}\left[(cd^{\dagger} +\hat{c}^{\dagger}\hat{d}) + \xi(\hat{c}^{\dagger}\hat{d}^{\dagger} + \hat{c}\hat{d})\right] + \omega_{F}
\end{gather}
where we have defined a set of reduced variables
\begin{gather*}
    \omega_{A} = \frac{\omega_{0}}{2}\left[
    \left(\frac{1}{\mu_{x}}-\frac{\eta_{z}}{\omega_{0}}\right)(1+\mu_{x})   + \frac{\eta_{x}}{\omega_{0}}(1-\mu_{x})\right], \\
    %%%
    \omega_{B} = \frac{\omega_{0}}{2}\frac{1-\mu_{x}}{4}\left(\frac{1}{1+\mu_{x}}\frac{3+\mu_{x}}{\mu_{x}}+\frac{\eta_{z}}{\omega_{0}} +\right.\\ \nonumber
    \left.\frac{\eta_{x}}{\omega_{0}}\frac{3\mu_{x}^{2}-2\mu_{x}+3}{1-\mu_{x}^{2}}\right), \\
    %%%
    \omega_{C} = -\frac{\omega_{0}}{8}\frac{\eta_{y}}{\omega_{0}}(1+\mu_{x}), \;\;\; \\
    %%%
    \omega_{D} = - \frac{\sqrt{2}}{4}f_{\xi x}^{1/2}\sqrt{\omega\omega_{0}}\tilde{\omega}_{zx}\frac{1-\mu_{x}}{\sqrt{1+\mu_{x}}}, \;\;\; \\
    %%%
    \omega_{E} = \gamma\sqrt{\frac{1}{2}(1+\mu_{x})}, \\
    %%%
    \omega_{F} = -\frac{\omega_{0}}{2}\left\{\left[\frac{1-\mu^{2}_{x}}{2\mu_{x}} + \left(\mu_{x}-\frac{\eta_{z}}{2\omega_{0}}\right)\right]2j +\right.\\ \nonumber
    \left.\frac{1}{2}\left[ \frac{1}{\mu_{x}} - \frac{\Delta\eta_{zx}}{\omega_{0}}\right](1-\mu_{x})\right\}. 
\end{gather*}
%%%
Just as we did in the deformed normal phases, we will proceed to diagonalize the Hamiltonian in Eq.~\ref{Hamiltonian frequencies}, by transforming it into quadratures and then applying rotations that decoupled them. This transformation reads
%%%
\begin{gather}
    \hat{c} = \sqrt{\frac{\omega}{2}}\left( \hat{x}+ \frac{i}{\omega}\hat{p}_{x} \right),\;\;\; 
    %%%
%    \hat{c}^{\dagger} = \sqrt{\frac{\omega}{2}}\left( \hat{x} -\frac{i}{\omega}\hat{p}_{x} \right)  \\
    %%%
    \hat{d} = \sqrt{\frac{\omega_{A}}{2}}\left( \hat{y} + \frac{i}{\omega_{A}}\hat{p}_{y} \right),\;\;\;
    %%%
%    \hat{d}^{\dagger} = \sqrt{\frac{\omega_{A}}{2}}\left( \hat{y} -\frac{i}{\omega_{A}}\hat{p}_{y} \right), 
\end{gather}
%%%
so the Hamiltonian is expressed as
%%%
\begin{gather}
    \hat{H} = \frac{1}{2}\left( \omega^{2}\hat{x}^{2} + \omega_{A}^{2}\chi_{+} \hat{y}^{2} + \hat{p}_{x}^{2} + \chi_{-}\hat{p}_{y}^{2}  +\right. 
    \\ \nonumber
    \left.\sqrt{\omega\omega_{A}}\kappa_{+}\hat{x}\hat{y} + \frac{\kappa_{-}}{\sqrt{\omega\omega_{A}}}\hat{p}_{x}\hat{p}_{y} -\epsilon_{1}\right).
\end{gather}
where
\begin{gather}
\epsilon_{1}=\omega_{0}\left[\frac{1-\mu^{2}_{x}}{2\mu_{x}} + \left(\mu_{x}-\frac{\eta_{z}}{2\omega_{0}}\right)\right]2j + 
\\ \nonumber
\omega_{0}\left[\left(\frac{1}{\mu_{x}}-\frac{\eta_{z}}{\omega_{0}}\right) + \frac{\eta_{x}}{\omega_{0}}(1-\mu_{x})\right]-\omega 
\end{gather}
%%%
Next, we consider the following variable changes
%%%
\begin{gather}
    \chi_{+} = \left(1+4\frac{\omega_{B}}{\omega_{A}}\right) ,\,\,\,  \\
    %%%
\kappa_{+} = 4\omega_{D} + \sqrt{2}f_{\xi x}^{1/2}\sqrt{\omega\omega_{0}}\tilde{\omega}_{zx}\sqrt{1+\mu_{x}},\\
    %%%
    \chi_{-} = \left(1-4\frac{\omega_{C}}{\omega_{A}}\right) , \,\,\, \\
    %%%
    \kappa_{-} = \sqrt{2}g_{\xi y}^{1/2}\sqrt{\omega\omega_{0}}\tilde{\omega}_{zy}\sqrt{1+\mu_{x}}.
\end{gather} 
Again, we apply two rotations: $\phi_{1}$ for $x$ and $y$, and $\phi_{2}$ for $p_{x}$ and $p_{y}$.
%%%
\begin{gather}
\left(
\begin{array}{c}
\hat{x} \\
\hat{y}  
\end{array}
\right)
\left(
\begin{array}{cc}
\cos{\phi_{1}} & \sin{\phi_{1}} \\
-\sin{\phi_{1}} & \cos{\phi_{1}}
\end{array}
\right)
\left(
\begin{array}{c}
\hat{q}_{1} \\
\hat{q}_{2}  
\end{array}
\right),\;\;\;
\\ \nonumber
\left(
\begin{array}{c}
\hat{p}_{x} \\
\hat{p}_{y}  
\end{array}
\right)
\left(
\begin{array}{cc}
\cos{\phi_{2}} & \sin{\phi_{2}} \\
-\sin{\phi_{2}} & \cos{\phi_{2}}
\end{array}
\right)
\left(
\begin{array}{c}
\hat{p}_{1} \\
\hat{p}_{2}  
\end{array}
\right).
\end{gather}
%%%
We obtain the following conditions for the angles $\phi_{1}$ and $\phi_{2}$ from eliminating the crossed terms $\hat{q}{1}\hat{q}{2}$ and $\hat{p}{1}\hat{p}{2}$,
%%%
\begin{gather}
    \tan2\phi_{1} = \frac{\sqrt{\omega\omega_{A}}\kappa_{+}}{\omega_{A}^{2}\chi_{+}-\omega^{2}}\;\;\; \text{and} \;\;\; \\ \nonumber \tan 2\phi_{2} =  \frac{\kappa_{-}}{\sqrt{\omega\omega_{A}}(\chi_{-}-1)}.
\end{gather}
%%%
The Hamiltonian reduces to:
%%%
\begin{gather}
    \hat{H} = \frac{1}{2}\left[(\epsilon_{1-}^{\text{S}}\hat{q}_{1})^{2}  + (\epsilon_{1+}^{\text{S}}\hat{q}_{2})^{2} + \right.
    \\ \nonumber
    \left.(\epsilon_{2-}^{\text{S}}\hat{p}_{1})^{2}  + (\epsilon_{2+}^{\text{S}}\hat{p}_{2})^{2} -\epsilon_{1}\right],
\end{gather}
%%%
with energies given by
%%%
\begin{gather}\label{epsilon 1S pm}
    \epsilon_{1\pm}^{\text{S}} = \sqrt{\frac{1}{2}\left((\omega^{2} + \omega_{A}^{2}\chi_{+})\pm \sqrt{(\omega_{A}^{2}\chi_{+}-\omega^{2})^{2} + \omega\omega_{A}\kappa_{+}^{2}} \right)}, \\
    %%%
    \epsilon_{2\pm}^{\text{S}} = \sqrt{\frac{1}{2}\left( (1+\chi_{-}) \pm \sqrt{(\chi_{-}-1)^{2} + \frac{\kappa_{-}^{2}}{\omega\omega_{A}}} \right)}.
\end{gather}
%%%
We decouple the Hamiltonian using quadratures.
%%%
\begin{gather*}
    \hat{q}_{1} = \frac{(\hat{a}^{\dagger}_{1} + \hat{a}_{1})}{\sqrt{2\omega_{\text{S}_{-}}}}, \;\;\; \hat{p}_{1} = i\sqrt{\frac{\omega_{\text{S}_{-}}}{2}}(\hat{a}^{\dagger}_{1} - \hat{a}_{1}), \\
    %%%
    \hat{q}_{2} = \frac{(\hat{a}^{\dagger}_{2} + \hat{a}_{2})}{\sqrt{2\omega_{\text{S}_{+}}}}, \;\;\; \hat{p}_{2} = i\sqrt{\frac{\omega_{\text{S}_{+}}}{2}}(\hat{a}^{\dagger}_{2} - \hat{a}_{2}), 
\end{gather*}
%%%
where, as in the deformed normal  case, $\omega_{\text{S}_{-}}=\epsilon_{1-}^{\text{S}}/\epsilon_{2-}^{\text{S}}$, $\omega_{\text{S}_{+}} =\epsilon_{1+}^{\text{S}}/\epsilon_{2+}^{\text{S}}$. Finally, the Hamiltonian reads in second quantized form as
%%%
\begin{gather}
    H =\epsilon_{0}^{\text{S}}+  \epsilon_{-}^{\text{S}}\hat{a}_{1}^{\dagger}\hat{a}_{1} + \epsilon_{+}^{\text{S}}\hat{a}_{2}^{\dagger}\hat{a}_{2} 
\end{gather}
%%%
with 
\begin{gather}
\epsilon_{0}^{\text{S}}=
\frac{1}{2}\left( \epsilon_{-}^{\text{S}} + \epsilon_{+}^{\text{S}}  -\epsilon_{1}\right),
\end{gather}
with the phase $\epsilon_{-}^{\text{S}}$ and amplitude $\epsilon_{+}^{\text{S}}$ modes being
%%%
\begin{gather}
\label{Modos de vibración Superradiante}
    \epsilon_{-}^{\text{S}} = \epsilon_{2-}^{\text{S}}\epsilon_{1-}^{\text{S}}, \;\;\; \epsilon_{+}^{\text{S}} = \epsilon_{2+}^{\text{S}}\epsilon_{1+}^{\text{S}}.
\end{gather}

%%%%%%%%%%%%%%%%%%%%%%%%%%%%%%%%%%%%%%%%%%
%%%%%%%%%%%%%%%%%%%%%%%%%%%%%%%%%%%%%%%%%%
\subsection{Deformed phase}
%%%%%%%%%%%%%%%%%%%%%%%%%%%%%%%%%%%%%%%%%%
%%%%%%%%%%%%%%%%%%%%%%%%%%%%%%%%%%%%%%%%%%

Here, we comment on the deformed phase in the Dicke limit ($\xi=1$) that suppresses the superradiant-$y$ phase~\cite{Rodriguez2018,Herrera2022}. It exists independently of the spin-boson coupling $\gamma$, while $\Delta\eta_{zy}\leq\omega_{0}$, and it is characterized by zero expectation value of the photon number but is not a normal phase. This can be recognized in the energy surface by a $\pi$ rotation of the extreme points. If we take $\xi=1$ in Eqs.~\ref{eq:eps1Npm} and~\ref{epsilon 1S pm}, $g_{\xi y}\rightarrow 0$, and  we get
%%%
\begin{gather}
\epsilon_{2\pm}^{N} = \sqrt{-\frac{\Delta\eta_{zy}}{2\omega_{0}}(1\pm 1)}, \;\;\;
    %%%     
\epsilon_{2\pm}^{S} = \sqrt{-\frac{\Delta\eta_{zy}}{2\omega_{0}}\mu_{x}(1\pm 1)}.
\end{gather}
Because $\epsilon_{\pm}^{\text{S},\text{N}}=\epsilon_{2\pm}^{\text{S},\text{N}}\epsilon_{1\pm}^{\text{S},\text{N}}$, the phase mode becomes vanishing, effectively suppressed, and the amplitude mode is imaginary for $\Delta\eta_{zx}\geq 1$, so it is not defined in the deformed phase. 

%%%%%%%%%%%%%%%%%%%%%%%%%%%%%%%%%%%%%%%%%%
%%%%%%%%%%%%%%%%%%%%%%%%%%%%%%%%%%%%%%%%%%
\section{Role of matter-matter interactions}
\label{sec:4}
%%%%%%%%%%%%%%%%%%%%%%%%%%%%%%%%%%%%%%%%%%
%%%%%%%%%%%%%%%%%%%%%%%%%%%%%%%%%%%%%%%%%%

%%%%%
\begin{figure*}[t]
\begin{center}
\begin{tabular}{c}
\includegraphics[width=0.9\textwidth]{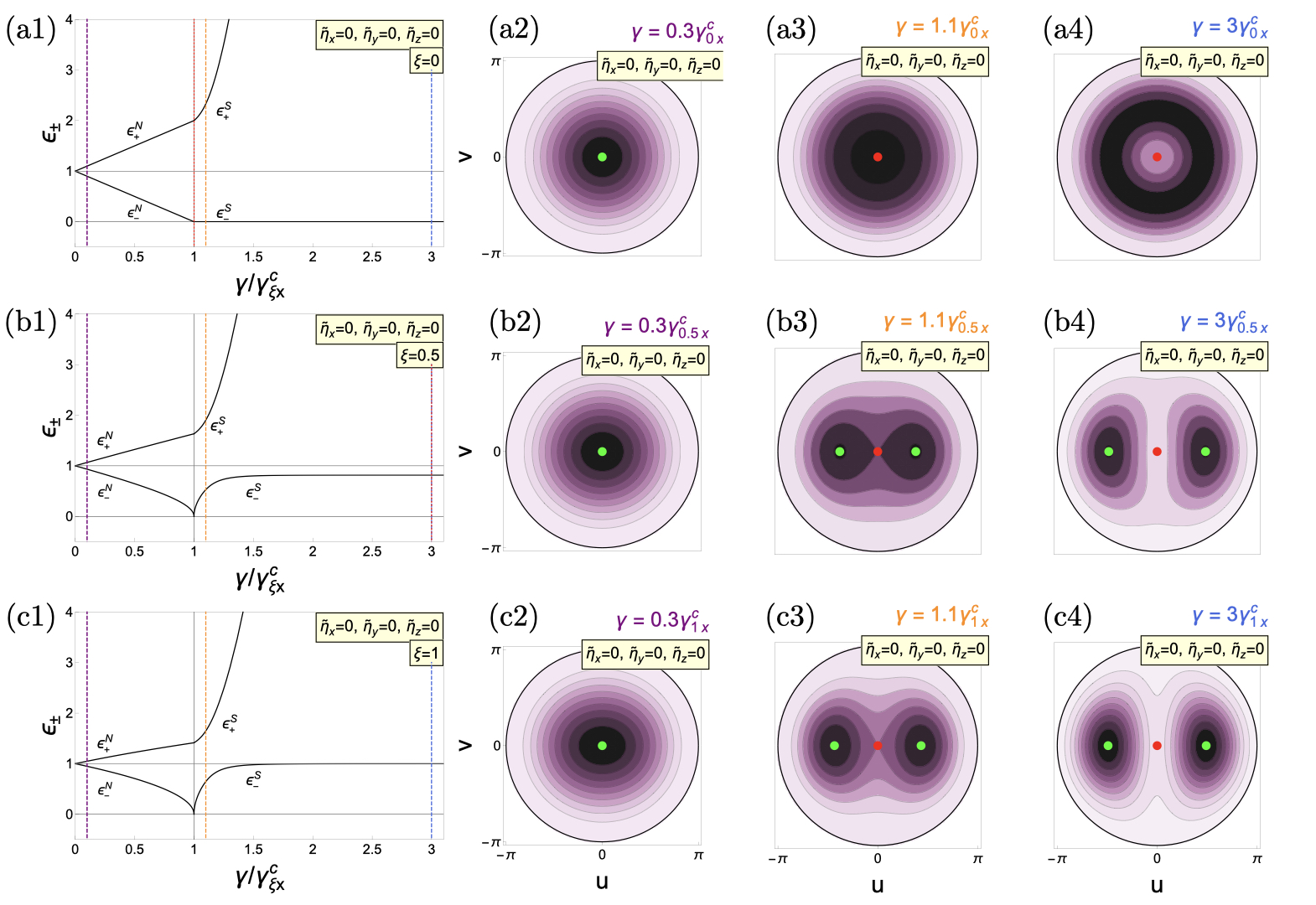}
\end{tabular}
\end{center}
\vspace{-20pt}
\caption{Polariton modes of the anisotropic Dicke model without material collective interactions. (a1) TC limit ($\xi=0$), (b1) Dicke limit ($\xi=1$), and (c1) anisotropic case ($\xi=0.5$). The critical coupling $\gamma_{\xi x}^{c}$ ($\gamma_{\xi y}^{c}$) is indicated by the vertical solid black (dotted red) line. Figs. (a2-a4), (b2-b4), and (c2-c4) depict the corresponding energy surfaces for the respective cases. The vertical dashed purple line shows the position of energy surfaces in the energy spectrum in the normal phases (a2), (b2), and (c2). The yellow one indicates the location of energy surfaces in the superradiant phase (a3),(b3), and (c3), while the blue line represents higher values of light-matter couplings (a4),(b4), and (c4). Green points in the energy surfaces represent energy minima, red ones indicate maxima and yellow points denote saddle points. Tilde variables are scaled to $\omega_{0}$. All cases are calculated in resonance  ($\omega=\omega_{0}=1$).}
\label{fig:ni} 
\end{figure*}
%%%%%

Next, we explore the behavior of the phase and amplitude modes as a function of the matter-matter interaction parameters for the Dicke ($\xi=1$), TC ($\xi=0$) and an anisotropic case, particularly $\xi=0.5$ as an example.

%%%%%%%%%%%%%%%%%%%%%%%%%%%%%%%%%%%%%%%%%%
\subsection{Absence of matter-matter interactions}
%%%%%%%%%%%%%%%%%%%%%%%%%%%%%%%%%%%%%%%%%%

As shown in Fig.~\ref{fig:ni}, first, we revisit the polariton modes without matter interactions. In Fig.~\ref{fig:ni} (a1) we show the polariton modes in the Tavis-Cummings limit. There, the lower curve corresponds to the phase mode $\epsilon_{-}^{N}$, while the upper one to the amplitude mode $\epsilon_{+}^{N}$. In the normal phase, the parametric evolution of both modes increases proportional to the light-matter coupling. Their energy gap is just the Rabi splitting $2\gamma$. This agrees with the overall behavior of the energy surface, which adopts a spherical well shape, a reflection of the conserved $U(1)$ symmetry [see Fig.~\ref{fig:ni} (a2)]. The dashed, vertical line marks the critical coupling $\gamma_{0x}^{(c)}=\gamma_{0+}^{(c)}$, separating the normal from the superradiant phases. There, the energy surface experiences a sudden change, taking the shape of a Mexican hat (see [Fig.~\ref{fig:ni} (a3-a4)]. Because of the onset of degeneration, we have spontaneous symmetry breaking. Once we enter the superradiant phase, the phase mode $\epsilon_{-}^{S}$ becomes a Goldstone mode with zero energy, depicted on the energy surface as the minimum energy ring of the Mexican hat~\cite{Fan2014,Hwang2016,Leonard2017,Chiacchio2018,Huang2023,Deng2023}. Instead, the amplitude mode $\epsilon_{+}^{S}$ depends on the brim of the Mexican hat. As it grows, it requires increasingly higher energy costs than the phase modes, which remain constant. Note that, in this case $\gamma_{0x}=\gamma_{0 y}=\gamma_{0-}$, so there is no presence of the superradiant-$(-)$ effects. In polariton terms, the two modes represent a maximal light-matter hybridization in resonance ($\omega=\omega_{0}$). The amplitude mode, being massive, can be identified as an upper-polariton branch that contains the most matter content, while the phase mode, being massless, becomes the lower-polariton branch with the most photonic content. Here, entering the superradiant phase implies a vanishment of the lower-polariton energy. 

The energy spectra on Fig.~\ref{fig:ni} (b1) and (c1) correspond to the standard Dicke model $(\xi=1)$ and the anisotropic case when $\xi=0.5$, respectively. Both cases exhibit quite similar behavior. In the normal phase, the behavior resembles the TC limit because the system is in the strong-interacting regime. At the critical coupling $\gamma_{\xi x}^{c}=\gamma_{\xi +}^{c}$, however, the phase mode tends to zero, while the amplitude mode to a fixed value given by 
\begin{gather}
\epsilon_{+}^{c}=\left(\frac{1}{2}\left(\omega^{2}+\omega_{0}^{2}\tilde{\omega}_{zx}^{2}\right)\right.\\ \nonumber
\left.\left[(1+\tilde{\omega}_{zy}^{2})+\sqrt{(1-\tilde{\omega}_{zy}^{2})^{2}+4\tilde{\omega}_{zx}^{2}\frac{(1-\xi)^{2}}{(1+\xi)^{2}}}\right]\right)^{1/2},
\end{gather}
which in the absence of interactions becomes
\begin{gather}
\epsilon_{+}^{c}=\sqrt{\left(\omega^{2}+\omega_{0}^{2}\right)\left(1+\left|\frac{1-\xi}{1+\xi}\right|\right)},
\end{gather}
and for the standard Dicke model is $\epsilon_{+}^{c}=(\omega^{2}+\omega^{2}_{0})^{1/2}$~\cite{Emary2003b}. When one passes the QPT, the phase mode gains a finite energy, becoming the rotonic mode~\cite{Fan2014,Chiacchio2018,Huang2023}, which converges to the finite value
\begin{gather} \label{eq:o}
\epsilon_{-}^{\gamma\gg 1}=\omega\sqrt{1+\left|\frac{1-\xi}{1+\xi}\right|},
\end{gather}
for larger couplings. This value comes from the energy cost to pass between the two degenerate minima (whose phases are $0$ and $\pi$) observed in Figs.~\ref{fig:ni} (b3) and (b4). The amplitude mode behaves quite similarly to the TC since the center of the energy surface presents an unstable point and tends quadratically as a function of $\gamma$, $\epsilon_{+}^{\gamma\gg 1}=(1+\xi)^{2}\gamma^{2}/\omega$ in the absence of matter interactions, which has the value $4\gamma^{2}/\omega$ in the standard Dicke model~\cite{Emary2003b}. 

It has been shown that increasing the anisotropy decreases the amplitude mode gap till one obtains the Goldstone mode. This can be seen from Eq.~\ref{eq:o}, where the limit $\xi\rightarrow 0$ leads to $\omega$ returning to the non-interacting value of photon field for the isotropic model~\cite{Emary2003b}. This critical feature has been proposed as an experimental signature for detecting the Goldstone mode~\cite{Baksic2014,Leonard2017}. We observe this feature in Fig.~\ref{fig:ni} (c1), and can be explained in terms of the change of the energy surface that leads to a new set of extremal points related to the superradiant-$(-)$ phase that reduces the energy of the phase mode~\cite{Herrera2022}. In this case, the presence of the counter-rotating terms that don't conserve the excitation number is reflected in the finite energy of the phase mode as a lower-polariton mode where light gains ``mass" close to the QPT and then becomes a massless but highly correlated state. In polariton terms, the lower-polariton energy becomes gaped. We notice, however, that the criticality of these cases is not affected by the superradiant-$(-)$ extreme points because, in the absence of interactions, they cannot be shifted, as it happens below. For the Dicke model $\gamma_{1 -}^{c}\rightarrow\infty$, and for the anisotropic case with $\xi=0.5$, the critical coupling $\gamma_{0.5 -}^{c}=3\gamma_{0.5 +}^{c}$ is deep in the superradiant phase and not shown in Fig.~\ref{fig:ni} (c1). 

%%%%%%%%%%%%%%%%%%%%%%%%%%%%%%%%%%%%%%%%%%
\subsection{Presence of matter-matter interactions}
%%%%%%%%%%%%%%%%%%%%%%%%%%%%%%%%%%%%%%%%%%

Next, we explore the effect of matter interactions over the polariton modes. In general, the $z$-interactions alone have the role of shifting both the critical coupling and the energy spectrum. To simplify the analysis we will focus on the role of the $x$ and $y$ interactions via the relevant parameters $\Delta\eta_{zx}$ and $\Delta\eta_{zy}$, respectively. 

\begin{figure*}[t]
\begin{center}
\begin{tabular}{c}
\includegraphics[width=0.9\textwidth]{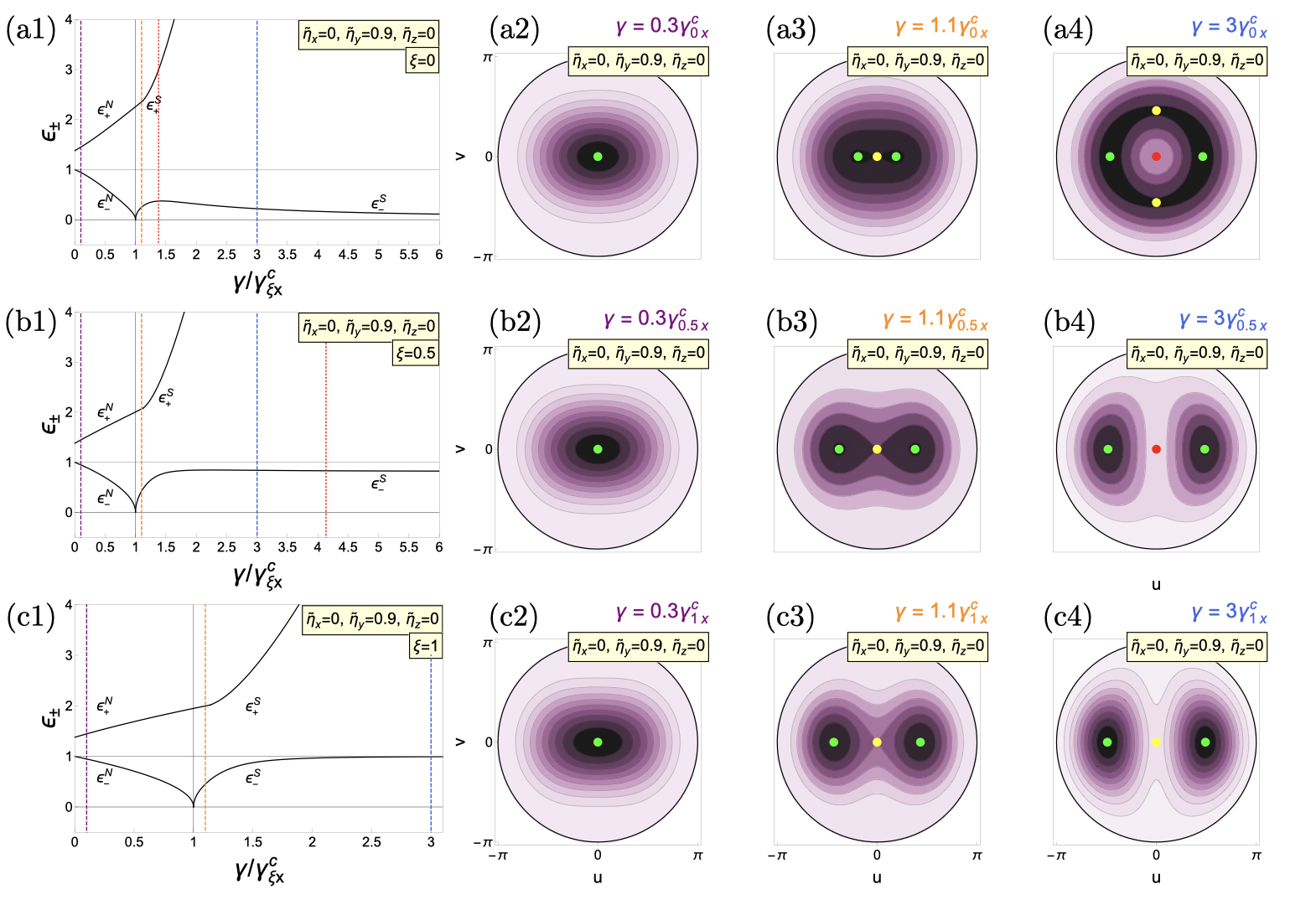}
\end{tabular}
\end{center}
\vspace{-20pt}
\caption{Same as Fig.~\ref{fig:ni} but with material collective interactions at $\eta_{y}=0.9\omega_{0}$ ($\Delta\eta_{zx}=0$, $\Delta\eta_{zy}=-0.9\omega_{0}$).}
\label{fig:ny} 
\end{figure*}

In Fig.~\ref{fig:ny} we show a representative value of the interactions given by $\eta_{y}=0.9\omega_{0}$ ($\Delta\eta_{zy}=-0.9\omega_{0}$), keeping $\eta_{x}=0$. Because of the interactions, in the normal phase, the energy surface widens along the $u$ direction for all $\xi$, so we call it the {\it deformed normal} phase~\cite{Herrera2022} [See Figs.~\ref{fig:ny} (a2), (b2), and (c2)]. In this case, the deformation does not substantially affect the phase and amplitude modes in the normal phase, except for an energy gap of the amplitude mode for zero spin-boson coupling given by
\begin{gather}
\epsilon_{+}^{N}=\omega_{0}\tilde{\omega}_{zx}\tilde{\omega}_{zy}.
\end{gather}
and representing the cost to climb the well in presence of interactions. Next, we cross the superradiant QPT. For the three cases under study ($\xi=0,0.5,1$), the amplitude mode is not significantly changed by the interactions compared to the case without. However, this is not the case for the phase mode in the TC limit ($\xi=0$). As the system enters the superradiant phase, the excitation number is not conserved anymore because of the deformation, so we do not have the Mexican hat potential anymore. As shown in Fig.~\ref{fig:ny} (a1), the phase mode gains energy, becoming rotonic. As we increase coupling, the onset of new extreme points at higher energies contributes to somehow restoring the original symmetry of the Mexican hat potential in the absence of interactions and reducing the mode's energy. Thus, as the spin-boson coupling increases, the phase mode slowly tends to a Goldstone-like phase mode. As a result, interactions produce a ``variable" mass roton mode. In both the anisotropic and Dicke limit cases, the results are similar to the case without interactions. After transitioning to the superradiant phase, two stable degenerate minimum points form, so there is no mechanism capable of decreasing the energy of the phase mode and eventually converging to a general energy value given by Eq.~\ref{eq:o}.

\begin{figure*}[t]
\begin{center}
\begin{tabular}{c}
\includegraphics[width=0.9\textwidth]{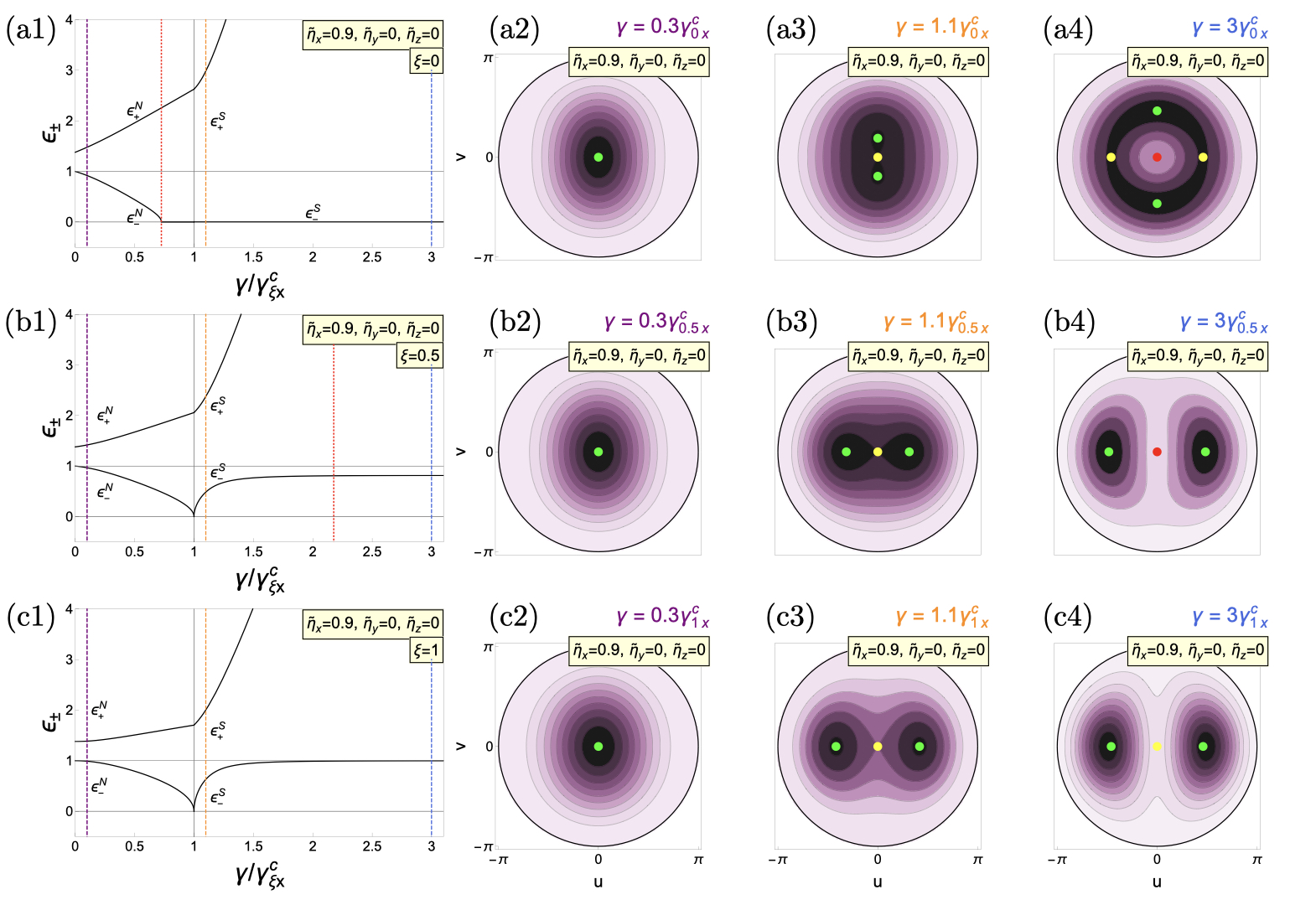}
\end{tabular}
\end{center}
\vspace{-20pt}
\caption{Same as Fig.~\ref{fig:ni} but with material collective interactions at $\eta_{x}=0.9\omega_{0}$ ($\Delta\eta_{zy}=0$, $\Delta\eta_{zx}=-0.9\omega_{0}$).}
\label{fig:nx} 
\end{figure*}

The situation is different for $x$-interactions, as shown in Fig.~\ref{fig:ny} for a representative value $\eta_{x}=0.9$ ($\Delta\eta_{zx}=-0.9$). We observe similar dynamics for the upper and lower-polariton modes as in the previous case. However, new phenomena emerge from the interplay between the anisotropy and matter interactions, i.e., the ability of the latter to shift the position of the superradiant-$(-)$ phenomena. Again, the normal phase becomes deformed for every $\xi$, but now with a different orientation, alongside the $v$ direction corresponding to the change from $y$- to $x$-interactions~\cite{Herrera2022} [See Figs.~\ref{fig:nx} (a2), (b2), and (c2)]. While the overall behavior of the amplitude mode remains the same, in the TC limit [Fig.~\ref{fig:nx} (a1)], we observe that the phase mode vanishes in the normal phase before the onset of the critical coupling. This is not a Goldstone mode but a suppression due to matter-matter interactions. As it can be seen from Eqs.~\ref{eq:eps1Npm}, two conditions make the phase mode vanish, either $f_{\xi x}=1$ or $g_{\xi y}=1$. As it can be shown in Fig.~\ref{fig:nx} (a1), the relative position of the critical coupling $\gamma_{0 x}^{c}$ with respect to $\gamma_{0 y}^{c}$ is shifted thanks to $\eta_{x}\neq 0$, creating a coupling regime where the phase mode cannot exist, although the amplitude mode does. This can be seen from the relation between the two critical couplings
\begin{gather}
\Delta\gamma=\gamma_{\xi x}^{c}-\gamma_{\xi y}^{c}=\gamma_{\xi x}^{c}\left(1-\frac{\tilde{\omega}_{zy}}{\tilde{\omega}_{zx}}\frac{1+\xi}{1-\xi}\right).
\end{gather}
and it occurs when $\Delta\gamma<0$. This unique effect results from the influence of the matter interactions over the competence between the superradiant-$+$ and superradiant-$y$ phases resulting from the anisotropy. This behavior does not occur in the Dicke and anisotropic cases [See Figs.~\ref{fig:nx} (a1) and (c1)], provided the onset of the superradiant-$y$ behavior occurs deep in the superradiant-$x$ one. However, for the anisotropic case, the interplay between the anisotropy and the matter interactions might shift the critical coupling $\gamma_{\xi y}^{c}=$ to suppress the mode. Beyond that, the Dicke and anisotropic cases, the behavior of the polariton modes remain qualitatively similar to previous cases, except for the $\pi$ rotation of the energy surfaces as one crosses the superradiant QPT, as it can be seen in Figs.~\ref{fig:nx} (b2) and (b3), as well as in Figs.~\ref{fig:nx} (c2) and (c3). 

%%%
%%%
%%%

%%%%%%%%%%%%%%%%%%%%%%%%%%%%%%%%%%%%%%%%%%
%%%%%%%%%%%%%%%%%%%%%%%%%%%%%%%%%%%%%%%%%%
\section{Geometric phase in presence of matter interactions}
\label{sec:5}
%%%%%%%%%%%%%%%%%%%%%%%%%%%%%%%%%%%%%%%%%%
%%%%%%%%%%%%%%%%%%%%%%%%%%%%%%%%%%%%%%%%%%

Next, we investigate the influence of matter interactions over the Berry or geometric phase induced by the bosonic field and the collective pseudospin over both polariton modes. The geometric phase is a topological quantum phase that the eigenstates of a Hamiltonian acquire when one goes adiabatically over a close path in the system's parameter space, accounting for the Hilbert space geometry~\cite{Berry1984,Berry1985}. The geometric phase has already been studied for the Dicke model~\cite{Plastina2006,Chen2006b}, extensions~\cite{Guerra2020,Lu2022}, and under the presence of matter interactions in the $z$-direction~\cite{Li2013}. Mainly because it becomes singular at criticality, it has been shown to serve as a signature for the superradiant QPT. Also, matter interactions could modify its scaling behavior~\cite{Li2013}. 

First, for comparison, we obtain the approximated ground-state energy without excitations. According to Eq.~\ref{lineal hamiltonian}, the scaled energy reads
%%%
\begin{gather}\label{ground energy}
    \epsilon_{0}=E_{0}(\alpha,\beta)/2j=\omega\alpha^{2} + \omega_{0}\beta^{2} - \frac{\omega_{0}}{2} - \\ \nonumber
    \gamma\sqrt{k}2\alpha\beta(1+\xi) + \eta_{x}k\beta^{2} + \eta_{z}\left(\beta^{2}-\frac{1}{2}\right)^{2}.
\end{gather}
%%%
It agrees to that calculated using coherent states as trial states~\cite{Herrera2022}. Then, by substituting Eq.~\ref{alphabeta} into Eq.~\ref{ground energy}, we obtain the ground-state energy for both the normal and superradiant phases
%%%
\begin{gather}\label{Hartree-Bogoliubov}
    \epsilon_{0} = \langle\Psi_{0}|2jH_{0}|\Psi_{0}\rangle =\\ \nonumber -\frac{j\omega_{0}}{2} 
    \left\{\begin{array}{cc}
       1-\frac{\eta_{z}}{2\omega_{0}}  & \gamma < \gamma^{c}_{\xi x} \\
   \frac{1}{2}\left(\mu_{x} + \frac{1}{\mu_{x}}\right) - \frac{\eta_{z}}{2\omega_{0}}  & \gamma > \gamma^{c}_{\xi x} 
    \end{array}\right.
\end{gather}
where $|\Psi_{0}\rangle$ is the ground-state of the system, and $\mu_{x}(\Delta\eta_{zx},\xi)=\left(\Delta\eta_{zx}/\omega_{0}+f_{\xi +}\right)^{-1}$, with $f_{\xi+}=\gamma^{2}/\gamma_{\xi+}^{2}$. The ground-state energy signals the superradiant QPT modified by the anisotropy and matter interactions at $\gamma_{\xi x}^{c}$ ($\mu_{x}=1$).

To calculate the geometric phase, we follow the standard procedure~\cite{FuentesGuridi2002,Carollo2003,Carollo2004,Bose2003,Chen2006b,Carollo2020}, where one considers a unitary transformation producing a cyclic trajectory over the parameter space. It may be an arbitrary adiabatic circulation generated by the photon number~\cite{Guerra2020} or over the Bloch sphere~\cite{Bose2003}. First, let us consider an adiabatic circulation generated by the photon number
\begin{gather}
\hat{U}(\phi_{n})=\exp\left(-i\phi_{n}\hat{n}\right)=\exp\left(-i\phi_{n}\hat{a}^{\dagger}\hat{a}\right).
\end{gather}
It provides an additional phase to the bosonic annihilation (creation) operator $\hat{a}\rightarrow\hat{a}e^{i\phi_{n}}$ ($\hat{a}^{\dagger}\rightarrow\hat{a}^{\dagger}e^{-i\phi_{n}}$), effectively displacing them. Next, we introduce a time-dependent unitary transformation, where the parameter $\phi_{n}(t)=\omega_{n} t$ varies adiabatically in the interval $[0,2\pi)$ with angular frequency $\omega_{n}$. The geometric phase is calculated in terms of the circuit integral of the Berry connection
\begin{gather}
\Gamma_{n}=i\int_{0}^{2\pi}\,d\phi_{n}\,\langle \Psi(\phi_{n})|\frac{d}{d\phi_{n}}|\Psi(\phi_{n})\rangle=\\ \nonumber 2\pi\langle\Psi_{0}|\hat{n}|\Psi_{0}\rangle,
\end{gather}
where $\langle\Psi_{0}|\hat{n}|\Psi_{0}\rangle=|\alpha|^{2}$ is just the expectation value of the photon number within the HPA. Hence, the geometric phase reads
\begin{gather}\label{geometric phase}
    \Gamma_{n} = j\left\{ \begin{array}{cc}
       0  & \gamma < \gamma^{c}_{\xi x} \\
       \frac{\pi}{2}\left(\frac{\gamma}{\gamma^c_{\xi x}}\right)^{2}\frac{\omega_{zx}^{2}}{\omega_{0}}(1-\mu_{x}^{2})  & \gamma > \gamma^{c}_{\xi x}
    \end{array}\right.
\end{gather}
%%%
For $\xi=1$, we recover the results of the standard ~\cite{Chen2006b} and modified Dicke models~\cite{Li2013}. Similarly, if now we consider a circulation generated by the excited pseudospin population $\hat{J}_{z}-j\mathbb{I}$
\begin{gather}
\hat{V}(\phi_{m})=\exp\left[-i\phi_{m}\left(\hat{J}_{z}-j\mathbb{I}\right)\right]=\exp\left(-i\phi_{m}\hat{b}^{\dagger}\hat{b}\right).
\end{gather}
Like in the boson case, this shifts the bosonic annihilation (creation) operator $\hat{b}\rightarrow\hat{b}e^{i\phi_{m}}$ ($\hat{b}^{\dagger}\rightarrow\hat{b}^{\dagger}e^{-i\phi_{m}}$), effectively displacing them. Changing adiabatically in time allows us to calculate the geometric phase as
\begin{gather}
\Gamma_{m}=i\int_{0}^{2\pi}\,d\phi_{m}\,\langle \Psi(\phi_{m})|\frac{d}{d\phi_{m}}|\Psi(\phi_{m})\rangle=\\ \nonumber 2\pi\langle\Psi_{0}|\hat{J}_{z}-j\mathbb{I}|\Psi_{0}\rangle,
\end{gather}
where $\langle\Psi_{0}|\hat{J}_{z}-j\mathbb{I}|\Psi_{0}\rangle=|\beta|^{2}$, so
\begin{gather}\label{geometric phase}
    \Gamma_{m} = j\left\{ \begin{array}{cc}
       0  & \gamma < \gamma^{c}_{\xi x} \\
       \pi\left(1-\mu_{x}\right)   & \gamma > \gamma^{c}_{\xi x}
    \end{array}\right.
\end{gather}

\begin{figure}[H]
\begin{center}
\begin{tabular}{c}
\includegraphics[width=0.7\columnwidth]{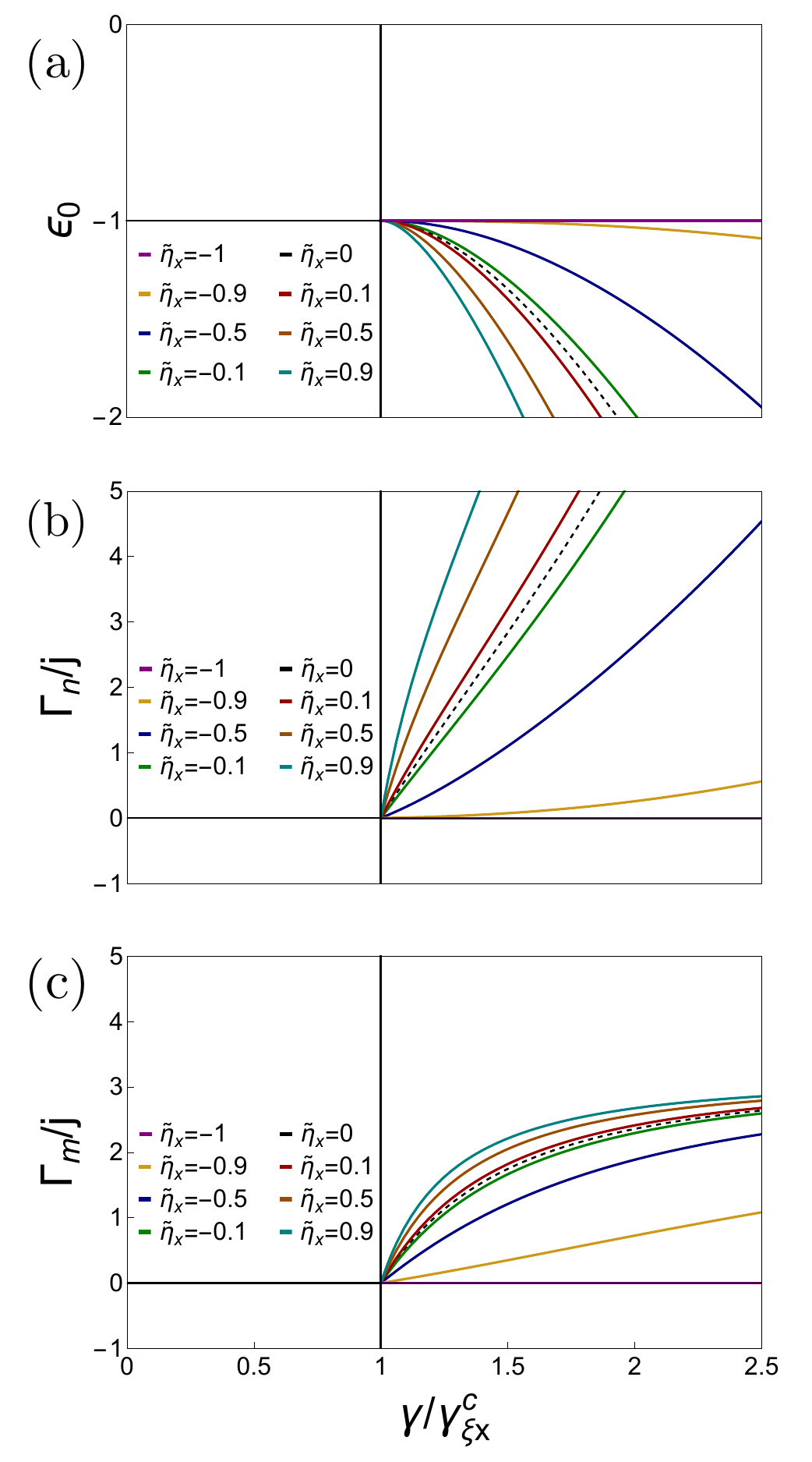}
\end{tabular}
\end{center}
\vspace{-20pt}
\caption{(a) Scaled ground-state energy $\epsilon_{0}=E_{0}/2j$ as a function of the light-matter coupling. (b) Scaled geometric phase $\Gamma_{n}$ considering an adiabatic circulation generated by the photon number. (c) Scaled geometric phase $\Gamma_{m}$ considering an adiabatic circulation generated by the collective pseudoespin $z$-projection. As indicated in the figures, we consider different values of the matter interactions $\Delta\eta_{zx}$ with $\eta_{z}=0$. The case with $\Delta\eta_{zx}=0$ is indicated as a dashed line. All cases are calculated in resonance $\omega=\omega_{0}=1$.}
\label{fig:geoph} 
\end{figure}

We notice that, for these contours, both the $\Gamma_{n}$ and $\Gamma_{m}$ geometric phases depend only in the ratio $f_{\xi x}$, as from Eq.~\label{eq:mu}, 
    $\mu_{x}^{-1} = \tilde{\omega}_{zx}/\omega_{0}\left(f_{\xi x}-1\right)+1$.
As a result, the behavior is qualitatively independent from $\xi$ and $\eta_{y}$. That it does not depend on $\eta_{y}$ is because the extreme points in the energy surface belonging to the superradiant-$y$ occur higher in energy, far away from the ground-state energy. Instead, the independence from $\xi$ is a feature that captures the universal behavior of the superradiant QPT. In Fig.~\ref{fig:geoph}, we show the results for both geometric phases $\Gamma_{n}$ and $\Gamma_{m}$and compare them to the scaled ground-state energy. We observe that the geometric phase identifies the position of the superradiant QPT, as expected. The geometric phases are sensitive to the sign of $\eta_{x}$ increasing with $\eta_{x}$.  Furthermore, we observe that the geometric phase also helps to identify the first-order phase transition occurring at $\Delta\eta_{x}=\omega_{0}$, where the geometric phase becomes zero independently of $\gamma$ in both cases and then turns negative. 

%%%%%%%%%%%%%%%%%%%%%%%%%%%%%%%%%%%%%%%%%%
%%%%%%%%%%%%%%%%%%%%%%%%%%%%%%%%%%%%%%%%%%
\section{Discussion and conclusions}
\label{sec:6}
%%%%%%%%%%%%%%%%%%%%%%%%%%%%%%%%%%%%%%%%%%
%%%%%%%%%%%%%%%%%%%%%%%%%%%%%%%%%%%%%%%%%%

We have derived analytic expressions for low-energy excitations of the anisotropic Dicke model in the presence of matter interactions by employing the Holstein-Primakoff transformation and neglecting higher-order terms in the thermodynamic limit. We explored the modes in the normal (ordered) phase, deformed by matter interactions, and the superradiant (ordered) phase from a unified point of view, whose interpretation is supported by the analysis of energy surfaces obtained using the corresponding classical Dicke model via coherent states~\cite{Herrera2022}.

The anisotropy in the Dicke model that, in this work, tunes the Hamiltonian between the Tavis-Cummings to the standard Dicke models produces a new set of extreme points and quantum phases, the superradiant-$(+)$ and $(-)$ phases, where superradiance is determined by the $(+)$ phase given that it lies lower in energy. Matter interactions serve two purposes: to shift the values of critical couplings and ground-state energy and to produce new extreme points in the energy surface, reflecting the onset of new quantum phases. Particularly, interactions in the $x$ and $y$ directions directly modify the criticality of the $(+)$ and $(-)$ phases, respectively, and shift their relative position in terms of the spin-boson coupling. 

%%%%% 
%%%%%

Beyond the shifting of critical couplings and ground-state energy associated to matter-matter interactons, we find unique effects from combining them with the anisotropy. In most cases, amplitude, massive, or upper-polariton mode behave qualitatively the same, just changing energetically to reflect the deformations of the energy surface, so interactions increase the gap to the phase mode, creating an energy cost given by $\omega_{0}\tilde{\omega}_{zx}\tilde{\omega}_{zy}$. Novel behavior appears mostly over the ground-state mode, i.e., the phase or lower-polariton mode. For the Dicke model, it becomes a roton mode that, for larger couplings, tends to the non-interacting photon energy $\omega$ and decreases in energy with the anisotropy. Without interactions it eventually transforms into the Goldstone mode. Instead, the TC limit exhibits a Goldstone mode in the absence of matter-matter interactions that becomes rotonic as the latter breaks the $U(2)$ symmetry, leaving conserved only the $Z_{2}$ one. 

A major result is that tuning the combination of matter interactions in the $z$ and $y$ directions can reduce the rotonic mode energy in the TC Hamiltonian, producing a ``variable" mass roton mode, or a lower-polariton that starts losing matter content with interactions. This is understood in terms of the energy surface, as the interactions influence on higher energies becomes less relevant to the ground-state, thus restoring to some extent the Goldstone-like behavior. Second, we discover that the Goldstone mode can be suppressed by tuning the interactions in the $z$ and $x$ directions by shifting the relative position between the critical couplings of the superradiant-$x$ and -$y$ phases. This suppression does not affect the amplitude mode. 

We anticipate these features could be detected and employed as an experimental signature to observe the Goldstone mode in novel setups, such as spin-magnon systems, where the interactions are tunable~\cite{Li2018,Bamba2022,MarquezPeraca2024}, as well as in others within the broad range of setups where the Dicke model can be realized, from quantum optics and atomic physics to condensed matter.

We also calculated the geometric phase for both boson and pseudospin contours, which equals the expectation value of the photon number and excited pseudospin in the phase mode within this approximation. We find that the role of the $z$ and $x$-interactions is to shift the phase. Still, it is independent of the $y$ interactions and the anisotropy parameter $\xi$, as it is a low-energy approach capable only of capturing the universal behavior of the superradiant-$x$ transition. Furthermore, the geometric phase becomes singular at the superradiant QPT, serving as a signature of criticality, where we recover previous results from Refs.~\cite{Chen2006,Li2013}. It also detects the first-order QPT that results from changing matter-matter interactions by changing sign. 

We conjecture that the predominance of the superradiant-$x$ phase will be inverted to the superradiant-$y$ phase in the case of the Anti-Tavis-Cummings model where now the rotating terms become smaller to the counter-rotating ones~\cite{AparicioAlcalde2011,Baksic2014,Shapiro2020}. Our model can be extended to explain this case but goes beyond the scope of this work. Finally, exploring the effect of the anisotropy and matter interactions over the polariton modes beyond mean-field is left for future work~\cite{Eastham2001,Boneberg2022}.

%%%%%%%%%%%%%%%%%%%%%%%%%%%%%%%%%%%%%%%%%%
\acknowledgments{R.H.R. acknowledges financial support from CONAHCyT postgraduate fellowship program. M.A.B.M. acknowledges financial support from the PEAPDI 2024 project from the DCBI UAM-I and from the PIPAIR 2024 project from the DAI UAM.
}
%%%%%%%%%%%%%%%%%%%%%%%%%%%%%%%%%%%%%%%%%%

%%%%%%%%%%%%%%%%%%%%%%%%%%%%%%%%%%%%%%%%%%
%\reftitle{References}
%\externalbibliography{yes}
\bibliography{references}
%%%%%%%%%%%%%%%%%%%%%%%%%%%%%%%%%%%%%%%%%%

%%%%%%%%%%%%%%%%%%%%%%%%%%%%%%%%%%%%%%%%%%
\end{document}